\documentclass[prb,aps,twocolumn]{revtex4}
\usepackage{graphicx}
\usepackage{subfigure}
\usepackage{bm}
\usepackage{amsmath}
\begin{document} 
\title{On dynamical tunneling and classical resonances}
\author{Srihari Keshavamurthy\footnote{Permanent address: Department of
Chemistry, Indian Institute of Technology, Kanpur, U.P. 208016, India.}}
\affiliation{Max-Planck-Instiut f\"{u}r Physik Komplexer Systeme, 
N\"{o}thnitzer Strasse 38,
D-01187 Dresden, Germany}
\date{\today}
\begin{abstract}
This work establishes a firm relationship between classical nonlinear
resonances and the phenomenon of dynamical tunneling. It is shown that
the classical phase space with its hierarchy of resonance islands 
completely characterizes dynamical tunneling and
explicit forms of the dynamical barriers can
be obtained only by identifying the key resonances. 
Relationship between the phase space viewpoint and the quantum mechanical
superexchange approach is discussed in near-integrable and
mixed regular-chaotic situations.
For near-integrable 
systems with sufficient anharmonicity the effect of multiple
resonances {\it i.e.,} resonance-assisted tunneling can be
incorporated approximately. 
It is also argued that
the, presumed, relation of avoided crossings to nonlinear resonances
does not have to be invoked in order to 
understand dynamical tunneling.
For molecules with low density of states the resonance-assisted
mechanism is expected to be dominant. 
\end{abstract}
\maketitle

\section{Introduction}
Intramolecular vibrational energy redistribution (IVR) 
in a molecule, from an initially prepared
nostationary state, is at the heart of chemical reaction dynamics.
The timescales and mechanisms involved in this process of energy flow
have been investigated in great 
detail by experiments and theory\cite{ivrrev}.
Although considerable progress has been made over the years there
are many aspects of this phenomenon whose understanding still eludes us.
Nevertheless it is expected that there is a part of the IVR mechanism
which can be usefully understood based on classical dynamics alone
and there is a part of IVR which is intrinsically quantum mechanical.
Both classical and quantum mechanisms coexist in a given molecule
and give rise to complicated spectral patterns and splittings.

It is now well established that a specific overtone excitation
in a molecule will undergo IVR if 
the initial nonstationary state, zeroth-order bright state (ZOBS), is
coupled to other 'dark' zeroth-order states via strong anharmonic resonances.
In such cases classical dynamics, at various
levels of detail and sophistication, 
can provide useful insights into 
the timescales and mechanisms of IVR\cite{clivr}.
On the other hand if the ZOBS is not close to any
resonance then classically one expects no IVR and hence
no fractionation of spectral lines. However, it is still possible
for IVR to occur via quantum routes. In other words the initial
state will mix with other states giving rise to spectral splittings
and, possibly, complicated eigenstates. 
Since the mixing is classically 'forbidden' it would be appropriate
to associate some kind of tunneling with such quantum routes
to energy flow.
Indeed such a suggestion was made over two decades ago\cite{davhel}
and the term 'dynamical tunneling' was coined. 
The notion of
tunneling is meaningless without one form of a barrier or other and
the term 'dynamical' is prefixed to distinguish from the usual
coordinate space tunneling through static potential barriers. 
The barriers in dynamical tunneling, in general,
are more subtle to identify and exist in the phase space.

Dynamical tunneling can have important consequences
for the interpretation of molecular spectra
since the fingerprints of IVR 
are spectrally encoded in the form of intensities and splittings. 
Traditionally dynamical tunneling
in the molecular context has been associated with the
occurence of local mode\cite{local} stretches in symmetric molecules. 
However it is important to emphasize that the concept of dynamical
tunneling is more general - any flow of quantum probability between
regions which are classically disconnected is dynamical tunneling.
The identification of classically disconnected regions requires
the knowledge of the phase space topology and hence it is
not very surprising that dynamical tunneling is inevitably linked
to the underlying classical dynamics.
In the context of local mode doublets in symmetric
systems important work by Jaff\'{e} and Brumer\cite{jafbrum}
and Kellman\cite{kell} 
provided classical phase space perspectives on the normal to local
transition. 
One of the reasons for
the interest in the near-degenerate local doublets has to do 
with their long lifetimes. Thus provided
one can experimentally prepare a high overtone, of say the OH-stretch
in H$_{2}$O, then the extremely small splitting provides a large window
of time to perform mode selective chemistry. The main problem with
such an idea is that
at such high vibrational excitations other Fermi resonances and 
rovibrational interactions could destroy the degeneracies\cite{quackarpc}.
We hope to shed some light on these issues in this work wherein
the system of interest does exhibit close degeneracies
despite multiple resonances and
a classical phase space which is mixed regular-chaotic.
Interestingly, similar issues have been addressed in the context
of the existence of discrete breathers in a network of nonlinear
oscillators\cite{db}.

One of the earliest
examples in symmetric systems are the local mode doublets observed in the
water molecule which were explained by Lawton and Child as due
to dynamical tunneling\cite{lawch}.
The experiments\cite{kerst}
of Kerstel {\it et.al.} on unusally slow (hundreds
of picoseconds) intramolecular
vibrational relaxation of CH-stretch excitation in (CH$_{3}$)$_{3}$CCCH
molecule has also been ascribed to dynamical tunneling by
Stuchebrukhov and Marcus\cite{stuma1}.
In this case the extremely slow IVR out of the CH-stretch is
an instance of dynamical tunnneling in nonsymmetric systems.
In a later experiment McIlroy and Nesbitt noted multiquantum state mixing
due to very small couplings (0.1 cm$^{-1}$ or less) in the CH stretch
excitation in propyne\cite{mcnes}. 
Recent experiments have
shown similar effects in the CH-stretch excitations
of triazine\cite{tri} and 1-butyne\cite{but} and 
there are reasons to expect dynamical tunneling to manifest in 
general rovibrational spectra\cite{mcd,kel,hart,lehkk,orti}.

Much of the theoretical understanding of dynamical tunneling
has come about from the analysis of bent triatomic ABA molecules.
The focus primarily has been on vibrations and this work is
no expception in this regard. However we note the important
work by Lehmann\cite{lehkk} wherein the nontrivial effects arising
due to coupling between local modes and rotations are
studied in detail.
Davis and Heller\cite{davhel} emphasized a phase space picture
and implicated classical resonances in the phase space as the agents
of dynamical tunneling. In this approach classical trajectories trapped
in one region of the phase space were imagined to be separated
by dynamical barriers, due to the resonances, from symmetrically 
equivalent regions of the phase space. The degeneracy is then 
broken by dynamical tunneling. 
A clear demonstration of the role of
isolated resonances was provied by Ozorio de Almeida as well\cite{ozo}.
A slightly different picture was 
provided by Sibert, Reinhardt and Hynes\cite{sihy}
in their work on energy flow and local mode splittings in the water
molecule. The setting was again in terms of classical nonlinear resonances
but the dynamical barrier was identified in angle space as opposed to
in action space. Later Hutchinson, Sibert and Hynes\cite{husihy} provided an
explanation for the quantum energy flow in terms of high
order perturbation theory. Stuchebrukhov and Marcus\cite{stuma2} reanalyzed
the ABA system in terms of chains of off-resonance virtual states
("vibrational superexchange") connecting any two degenerate local
mode states. An important result was the equivalence between
the vibrational superexchange approach and the usual WKB expression
for splittings in a doublewell system.
This equivalence between perturbative and the
WKB approaches was shown for a classically integrable case.
Demonstration of such an equivalence in more general
situations is an open problem.
Despite seemingly different approaches a crucial ingredient
to any description of dynamical tunneling involves
the various anharmonic resonances. 
Depending on the level of excitation the underlying phase space
can exhibit several (near) equivalent regions separated by nonlinear
resonances (near-integrable phase space) or chaos (mixed phase space).
For large molecules with sufficient density of
states, in the near-integrable regime,
Heller has conjectured\cite{hel9599} that a nominal
10$^{-1}$-10$^{-2}$ cm$^{-1}$ broadening of spectroscopically
prepared zeroth-order states is due to dynamical tunneling between
remote regions of phase space facilitated by distant resonances.

The aforementioned conjecture is based on the notion that
phase space is the correct setting for an understanding of
dynamical tunneling. 
It is thus natural to expect the splittings to be sensitive
to the structure of the phase space. It also follows that
given the phase space structure 
one ought
to be able to compute the associated splittings.
General forms for splittings cannot be written down easily since
explicit forms of the dynamical barriers
in the phase space, seperating quantum
states localized in distant regions of the phase space,
can only be provided upon identification
of the key nonlinear resonances.
This is also tanatamount to identifying the mechanism of dynamical
tunneling and hence IVR.
Two factors make this a difficult
problem. Firstly, our understanding of global structure
of multidimensional classical
phase space is still in its infancy. 
Thus without a global view of the phase space at energies corresponding
to the doublets it is difficult to identify the main nonlinear
resonances and perhaps the presence of appreciable stochastic layers.
Secondly, and related to the first factor,
important work over the
last decade has established
that dynamical tunneling is sensitive not only 
to the nonlinear resonances\cite{rat1,rat2}
but also to the classical stochasticity\cite{cat}. 
In the near-integrable regimes dynamical tunneling can be
enhanced by many orders of magnitude due to the various
nonlinear resonances. 
This is called as resonance-assisted tunneling
and it has been suggested\cite{elt} 
that the nonlinear resonances
play an important role in 
mixed phase space scenario as well.
In cases where an appreciable chaotic region separates the 
two symmetry related regular zones one has the phenomenon of
chaos-assisted tunneling. The hallmark of chaos-assisted tunneling
is the erratic fluctuations of the splittings\cite{cat} due to 'chaotic' states
having nonzero overlaps with the regular doublets. In particular
the splittings show algebraic dependence on $\hbar$ as opposed to
the integrable $\exp(-1/\hbar)$ scaling. 
Experiments\cite{expt} on a wide variety of systems have highlighted
the role of the phase space structures in dynamical tunneling.

Is the evidence for the sensitivity of dynamical tunneling
to the phase space structures already present in the
molecular spectra? The answer to this question, apart
from a fundamental viewpoint, is also potentially relevant
to mode-specific chemistry and control. 
In this context, it is significant to note
that most of the earlier works
in the molecular context have focused
on dynamical tunneling where only one resonance was involved with
no chaos. A notable exception
is the work of Davis and Heller\cite{davhel} wherein a hint to the role
played by classical chaos was provided.
The sensitivity of dynamical tunneling to the underlying
phase space structure can have important ramifications
in the molecular context. 
Phase space of molecular systems at higher energies, corresponding to
high overtone transitions, are typically a mixture of regular and
chaotic regions. The presence of stochasticity can lead to
the dynamical tunneling being enhanced (or suppressed) by several orders of
magnitude and hence highly mixed states and complicated
spectral patterns. 
Even in the near-integrable regimes tiny induced
resonances can arise as a result of the primary resonances which
can enhance or suppress dynamical tunneling between symmetry related
modes.    
Alternatively, molecule-field interactions can\cite{mfield} lead to
creation (destruction) of nonlinear resonances which can lead
to significant enhancement or supression of energy flow.

The purpose of this work is to establish the role of classical
resonances in the near-integrable phase space for dynamical tunneling 
and the nontrivial effect of higher order induced resonances and
perhaps chaos. 
We focus on the symmetric case in this work but
the role played by the nonlinear resonances is expected to hold
in the nonsymmetric situations as well.
In the near-integrable regime it is possible for more than one
nonlinear resonance zone to manifest in the phase space. In such
instances the single rotor picture\cite{ozo,sihy} is not sufficient to account
for dynamical tunneling. Yet it is shown
that hopping across the various resonance islands
is an approximate picture that  
yields qualitative (and perhaps quantitative)
insights into the splitting patterns.
We start by motivating the model Hamiltonian in
section~\ref{modham}. In section~\ref{qv}
the concept of dynamical tunneling
and quantum viewpoints are illustrated. The phase space viewpoint
and relations to the quantum viewpoint are discussed in section~\ref{psv0}
by focusing on specific cases. 
In section~\ref{psv1} two issues regarding avoided
crossings and influence of very small induced resonances
are addressed briefly.
Complications due to multiple resonances in the near-integrable limit
and the resulting resonance-assisted tunneling are illustrated
in section~\ref{psv2}. The possibility of chaos-assisted tunneling
occuring in the system is explored in section~\ref{psv3}.
A brief summary is provided in section~\ref{summ}.

\section{Model Hamiltonian}
\label{modham}
As discussed in the introduction
attempts to understand dynamical tunneling, from
a classical phase space viewpoint,
in three or more degrees of freedom systems is ambitious
without studying the detailed correspondence in lower
dimensions with multiple resonances. 
In this regard the effective spectroscopic Hamiltonian for water
due to Baggott\cite{bagg}
provides a good model system.
The Hamiltonian is given by
\begin{equation}
\hat{H}_{C}=\hat{H}_{0}+g' \hat{V}^{(12)}_{1:1}+\gamma \hat{V}^{(12)}_{2:2}
+\frac{\beta}{2\sqrt{2}}(\hat{V}^{(1b)}_{2:1}+\hat{V}^{(2b)}_{2:1})
\label{eq1}
\end{equation}
with, $g'=g+\lambda'(n_{1}+n_{2}+1)+\lambda''n_{b}$ and
\begin{eqnarray}
\hat{H}_{0}&=&\omega_{s}(n_{1}+n_{2})+\omega_{b}n_{b}
+x_{s}(n^{2}_{1}+n^{2}_{2}) \nonumber \\
&+&x_{b}n^{2}_{b}+x_{sb}n_{b}(n_{1}+n_{2})+
x_{ss}n_{1}n_{2}
\end{eqnarray}
describing the anharmonic local stretches $(1,2)$,
and bend $(b)$. 
The various parameter values are $\omega_{s} = 3885.57, \omega_{b}=1651.72,
x_{s} = -81.99, x_{b} = -18.91, x_{sb} = -19.12$ and, $x_{ss} = -12.17$
cm$^{-1}$.  
The $j^{th}$ mode occupancy is $n_{j}=a^{\dagger}_{j}a_{j}$
with $(a^{\dagger}_{j},a_{j})$ being the harmonic oscillator
creation and destruction operators.
The perturbations, anharmonic resonances,
\begin{equation}
\hat{V}^{(ij)}_{p:q}=(a^{\dagger}_{i})^{q}(a_{j})^{p}+
(a^{\dagger}_{j})^{p}(a_{i})^{q}
\end{equation}
connect zeroth-order states
$|{\bf n}\rangle$, $|{\bf n}'\rangle$ with $|n_{i}'-n_{i}|=q$ and
$|n_{j}'-n_{j}|=p$. 
The strengths of the resonances are $g=-56.48, \lambda'=3.02, \lambda''=-0.18,
\gamma = -0.91$, and $\beta=26.57$ cm$^{-1}$.
In order to perform detailed studies on the above Hamiltonian
in this work we also analyze subsystems obtained by
retaining specific resonances. These subsystems, denoted
by $A$ and $B$, are described
by the Hamiltonians $H_{A}=H_{0}+\beta(V^{(1b)}_{2:1}+V^{(2b)}_{2:1})$ and
$H_{B}=H_{A}+\gamma V^{(12)}_{2:2}$ respectively.
The full system with all the resonances will be denoted by $C$.
The reason for this specific choice of subsystems has to do with the
fact that classically $H_{A}$ and $H_{B}$ exhibit near-integrable
dynamics whereas $H_{C}$ has mixed regular-chaotic dynamics.
$\hat{H}$
is effectively two dimensional due to the existence of the  
conserved 
polyad $P=(n_{1}+n_{2})+n_{b}/2$ {\it i.e.,} $[\hat{H},\hat{P}]=0$.
Spectroscopic Hamiltonians can be 
obtained by a fit to the high resolution spectra
or from a perturbative analysis of an existing
high quality {\it ab initio} potential
energy surface\cite{joy}.
In any case such Hamiltonians, and variants thereof, have
proved quite useful in studying the highly excited vibrations
of many small molecules\cite{jacjpc}.
Couplings to rotations and large amplitude modes are certainly
important\cite{lehkk} but a detailed phase space analysis of such general
rovibrational Hamiltonians is beyond the scope of this
work.

The classical limit\cite{ksjcp} 
of the above Hamiltonian can be obtained by using the
Heisenberg correspondence, $\hat{a}_{j} \leftrightarrow \sqrt{I_{j}}
\exp(i \theta_{j})$, and results in 
a nonlinear multiresonant
Hamiltonian 
\begin{eqnarray}
H({\bf I},{\bm \theta})&=&H_{0}({\bf I}) \\
&+& g_{c}' \sqrt{I_{1}I_{2}} \cos(\theta_{1}-\theta_{2})+
\gamma_{c} I_{1} I_{2} \cos(2\theta_{1}-2\theta_{2}) \nonumber \\
&+& \beta_{c} I_{b} [\sqrt{I_{1}} \cos(\theta_{1}-2\theta_{b})+
\sqrt{I_{2}} \cos(\theta_{2}-2\theta_{b})] \nonumber 
\label{cham}
\end{eqnarray}
with $(I_{j},\theta_{j})$ corresponding to the action-angle variables
associated with the mode $j$. 
Note that the actions ${\bf I}$ and the zeroth-order quantum numbers
${\bf n}$ are related by the correspondence ${\bf I} \rightarrow
({\bf n}+1/2)\hbar$ for simple vibrations. 
The various resonant couplings are realted to the quantum coupling
strengths by $g_{c}'=2g'$, $\gamma_{c}=2 \gamma$, and $\beta_{c}=2 \beta$.
We note that another reason for the choice of this system has
to do with the fact that fairly detailed classical-quantum correspondence
studies of the highly excited eigenstates have been already
performed\cite{ksjcp,kelljcp}. 

\begin{figure}
\includegraphics*[width=3in,height=3in]{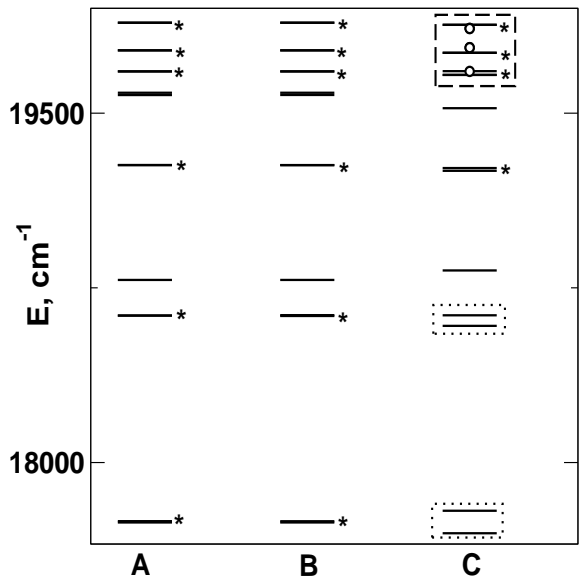}
\caption{A portion of the energy levels of the model
effective Hamiltonian corresponding to polyad $P=6$. The
full system is denoted by C and some of the experimental
energy levels are shown as open circles. B and A show
the energy levels for the subsystems with specific
resonances removed.
Some of the closely spaced
doublets are indicated by $*$. In C the box around the top three
levels indicates levels with $4,5$, and $6$ quanta of
excitation in the stretch mode.}
\label{fig1}
\end{figure}

In this study the focus is on a specific set of 
molecular parameters (provided above)
which are representative of the H$_{2}$O molecule\cite{bagg,ksjcp}. 
However, at the outset, we emphasize that this choice is by no means
special for the analysis and conclusions of this work. 
The goal of
this work is not to provide accurate quantitative estimates for
the splittings - something that can be trivially calculated quantum
mechanically for the model Hamiltonian. The emphasis here is on
establishing and understanding, qualitatively, the important role of nonlinear
resonances in dynamical tunneling and for this purpose the above model
Hamiltonian provides a good paradigm.   
In particular
we choose the states corresponding to the polyad $P=6$ which lie in
a energy range of $[16600,21300]$ cm$^{-1}$ above the
ground state. There are two main reasons for such a choice.
Firstly the motivation is to understand aspects of dynamical
tunneling for excited vibrational states both in the presence of
nonlinear resonances and chaos\cite{ksjcp}. 
Secondly the exhaustive compilation\cite{tenny}
of the experimental energy levels of H$_{2}$O by Tennyson {\it et.al.}
does show three doublets in this energy region. 
In local mode notation these doublets are $(60)^{\pm}0, (50)^{\pm}2$, and
$(51)^{\pm}0$. A few other local modes are also reported with only one
of the parity states like $(40)^{-}4$.
Although the
effective Hamiltonian used here is perhaps not the best it still
provides some correspondence to the patterns observed in the 
experimental results. 
Moreover at such high levels of excitation,
certainly for the bending mode,
there can be significant interpolyad mixings\cite{bykov}
which require analyzing a 
three degree of freedom system. Thus we use
the effective Hamiltonian as a toy model which would guide further,
more rigorous, studies in full dimensionality.
To give an idea about the validity of the effective Hamiltonian
in the chosen energy range, in Fig.~\ref{fig1} we show the energy levels
as compared to the experimental levels and the agreement is
satisfactory.

\section{Dynamical tunneling: illustration and quantum viewpoint}
\label{qv}

\begin{figure}
\includegraphics*[width=3in,height=2.0in]{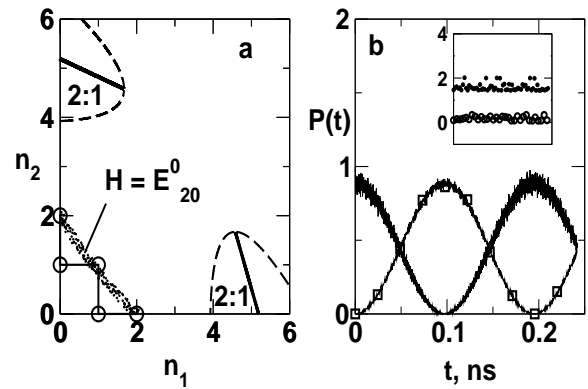}
\caption{Classical and quantum viewpoints on
dynamical tunneling for subsystem $A$.
The states are $|02\rangle$ and $|20\rangle$.
(a) The state space for polyad $P=6$ is shown with the regions
strongly influenced by the 2:1 anharmonic resonances. An
example of a minimal (solid)
superexchange chain connecting
the states
is shown.
The energy contour satisfying $H_{0} \approx E^{0}_{20}$
is shown connecting the two states.
(b) Quantum survival probabilties for $|20\rangle$
and $|02\rangle$ (squares) are shown for $0.25$ ns.
Note the almost
coherent transfer of population
with a period of $0.20$ ns.
The inset shows a classical trajectory initiated near $|20\rangle$ for
a time period of about $0.25$ ns. The time variations
of the actions $(I_{1}-1/2)$ and
$(I_{2}-1/2)$ are shown as filled and unfilled
circles respectively. The actions stay trapped for the
entire period.}
\label{fig2}
\end{figure}

In the rest of the paper the zeroth-order states will be denoted by
$|n_{1}n_{2}\rangle$ since the bend quantum number $n_{b}$ is fixed
by the polyad $P$.
In order to illustrate the concept
of dynamical tunneling
we consider  
zeroth-order degenerate states $|20\rangle$ and $|02\rangle$
with only the 2:1 resonances 
present {\it i.e.,} $g'=\gamma=0$ in eq.~\ref{eq1}. 
As $\langle 20|\hat{V}^{(1b)}_{2:1}|02\rangle = 0 =
\langle 20|\hat{V}^{(2b)}_{2:1}|02\rangle$ 
the state $|20\rangle$ has no direct coupling to the
symmetric counterpart $|02\rangle$. 
Nonetheless there is an indirect coupling via the 2:1 resonances.
However, as shown in Fig.~\ref{fig2}a the zeroth-order state is
quite far from the region of state space which can potentially
come under the influence of the 2:1 anharmonic resonances.
A classical trajectory initiated with initial conditions corresponding
to the state $|2,0\rangle$ will continue to stay in that region
of the phase space forever without reaching the symmetric
region corresponding to the state $|02\rangle$.
One such classical trajectory, for about $0.25$ ns,
in phase space is shown in Fig.~\ref{fig2}b (inset).
This classical trapping occurs despite no apparent energetic barriers
since in Fig.~\ref{fig2}a we show the classical actions
satisfying $H_{0} \approx E^{0}_{20}$ which connects the two states.
However, contrary to the classical observation,
in Fig.~\ref{fig2}b the quantum survival probability of the state
$|20\rangle$ shows transfer of population to $|02\rangle$ with a period
of about $0.20$ ns. In a two-state approximation this corresponds
to a splitting of the degenerate levels $\Delta_{2} \approx 0.17$ cm$^{-1}$.
{\em This is the phenomenon of dynamical tunneling.}  
Previous discussions\cite{sihy,husihy,stuma2} 
of dynamical tunneling have used the effective
Hamiltonian with only the primary 1:1 present. Here we delibrately
show this effect with the 2:1s only to emphasize the role
of induced resonances. 

A possible explanation
of the tunneling arises from the  
vibrational superexchange\cite{husihy,stuma2} perspective 
wherein zeroth-order 
states coupled locally by the 2:1 perturbations to
$|20\rangle$ and $|02\rangle$ are considered.
One then constructs all possible perturbative chains connecting the 
states $|20\rangle$ and $|02\rangle$. An example of such
a chain, shown in Fig.~\ref{fig2}a,
is $|20\rangle \rightarrow |10\rangle \rightarrow
|11\rangle \rightarrow |01\rangle \rightarrow |02\rangle$.
The contribution to the splitting is 
\begin{equation}
\beta^{4}\frac{\langle 20|\hat{V}^{(1b)}|10\rangle \langle
10|\hat{V}^{(2b)}|11\rangle \langle 11|\hat{V}^{(1b)}|
01\rangle \langle 01|\hat{V}^{(2b)}|02\rangle}{
(\Delta E^{0}_{10})^{2}(\Delta E^{0}_{11})}
\end{equation}
with $\Delta E^{0}_{n_{1}n_{2}} \equiv
E^{0}_{20}-E^{0}_{n_{1}n_{2}}$.
In principle there are an infinite number of chains that connect the
two degenerate states. In practice, due to the energy denominators
it is sufficient to consider the 
minimal length chains\cite{self}. 
By minimal length chains one means inclusion of all possible perturbative
terms at the lowest order of perturbation necessary. For instance in the
example above it is necessary to go to atleast fourth order
in perturbation theory, thus scaling
like $\beta^{4}$, to get a finite value for the splitting.
In our case
there are six minimal chains and summing
the contributions one obtains a splitting of
about $0.18$ cm$^{-1}$ which compares well with the exact splitting.

\begin{figure}
\includegraphics*[width=3in,height=3in]{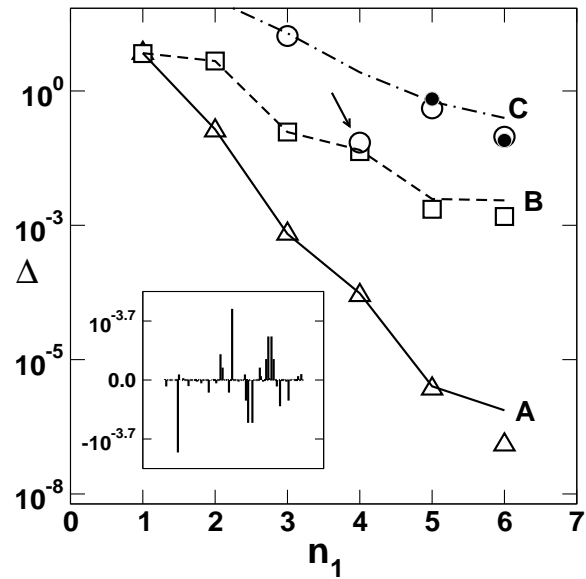}
\caption{Comparison of the exact quantum (symbols) dynamical
tunneling splittings $\Delta$, in
cm$^{-1}$, to the superexchange results (lines).
Shown are
the splitting between the states $|n_{1}=r,n_{2}=0\rangle$ and
$|0r\rangle$.
The $r=1,2$ cases are not shown for the full system
since on inclusion of the large 1:1 they are not local modes but resonant
normal modes with large splittings. The experimental values for the splitting
for $r=5,6$ are shown as filled circles for comparison. The arrow
indicates the substantial deviation of the superexchange result from the
exact value for $r=4$. The inset shows the contribution from $130$
terms for the superexchange calculation of $\Delta_{6}$ in case A.}
\label{fig3}
\end{figure}

More generally, splitting for any set of 
degenerate states $|r0\rangle$
and $|0r\rangle$,
at minimal order, can be calculated as\cite{self}:
\begin{equation}
\Delta_{r} = \sum_{abc} \beta^{a}g^{b}\gamma^{c} \sum_{\mu} \Delta_{r}
(\Gamma^{(\mu)}_{abc})
\label{mhopt}
\end{equation}
with $\mu$ being an index of all possible chains $\Gamma_{abc}$
for a particular choice of $a,b,c$ satisfying the constraint
$a+2b+4c=2r$. Indeed such a calculation can be done on our
model system and the resulting perturbative estimates 
are compared to the exact splittings in Fig.~\ref{fig3}. For further
discussions the superexchange calculations have been
compared to the exact results
for cases $A,B$, and $C$. 
At the outset note that the perturbative
calculations essentially reproduce, to quantitative accuracy, almost
all of the splittings. This degree of accuracy, given high orders of
perturbation and number of chains, is pleasantly surprising. For
instance the inset to Fig.~\ref{fig3} shows that $130$ contributions of
varying signs, some as large as $10^{-4}$, conspire to
yield the correct splitting of about $10^{-7}$ for $\Delta_{6}$.
Two important observations can be made at this juncture. First, the
significant role of the tiny 2:2 resonance is clear in that
addition of this 2:2 to the 2:1s enhances 
the splittings by many orders of magnitude - something which is
not readily apparent from Fig.~\ref{fig1}.
For even values of $r$, especially $r=2$, this is expected
since the 2:2 provides a dominant coupling. In fact for $r=2$
one expects $\Delta_{2} \approx 4 \gamma = 3.6$ cm$^{-1}$ 
which compares very well to the exact value of $3.8$ cm$^{-1}$.
However, it is significant to note that such a calculation
for $r=6$ is too small by a factor of five and that the odd $r$
states would not split at all in the presence of a 2:2 coupling alone.
Secondly in all three cases $A,B$, and $C$ the splittings more or less
monotonically decrease with increasing stretch excitations.
The one exception 
is in the full system (case $C$) for the state with $r=4$.
Later it is shown that this is an instance of suppression of tunneling
due to an avoided crossing between the doublet and a third
'intruder' state. Incidentally in the experimental data\cite{tenny} pertaining
to $r=4$ among the doublets, denoted in the local mode basis
as $(40)^{\pm}4$, only the state $(40)^{-}4$ 
is reported\cite{tenpers}. 
This, given
our model Hamiltonian, could be a pure accident!

It is important to note that the superexchange approach invokes the
2:1 resonant terms without any reference to the classical phase space.
This surprisingly good accuracy seems to hold whenever the
phase space is integrable or near-integrable and is perhaps related
to the equivalence of the superexchange to the WKB approach\cite{stuma2}.
If this observation is true in multidimensional systems then
a phase space perspective on dynamical tunneling should be
able to provide some insights. 
Apart from providing a phase space analog to the superexchange
approach it would also be possible to study the extent of
sensitvity of dynamical tunneling to the various classical structures.
The rest of the paper is dedicated to
uncovering precisely such a phase space picture.

\section{Dynamical tunneling: phase space viewpoint}
\label{psv0}
\subsection{Induced resonances and dynamical barriers}
\label{psv}

In what follows, the
classical Poincar\'{e} surface of sections
are plotted in 
$((I_{1}-I_{2})/2,(\theta_{1}-\theta_{2})) \equiv (K_{1}/2,2\phi_{1})$ at
constant values of the sectioning angle
$\phi=\theta_{1}+\theta_{2}-4\theta_{b}$ with $\dot{\phi} > 0$. In this
representation the normal-mode resonant regions appear around $I_{1} \sim
I_{2}$. Local modes regions correspond to above and below the
normal-mode regions and large 2:1 bend/local-stretch resonant islands
appear in the local mode regions.

\begin{figure}
\includegraphics*[width=3in,height=3in]{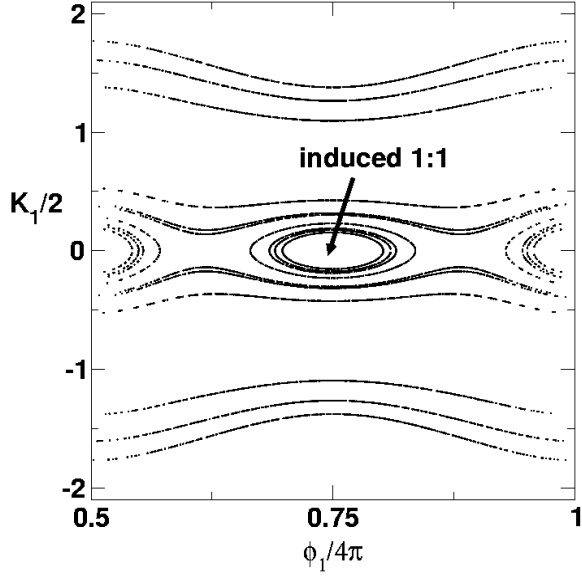}
\caption{Classical Poincar\'{e} surface of section for $H_{A}$ at
$E \approx E^{0}_{20}$ showing the induced 1:1 resonance island
due to the two stretch-bend 2:1 resonances.}
\label{fig4}
\end{figure}

We begin by asking the question as to what possible
structure in the classical phase space is mediating the
dynamical tunneling in Fig.~\ref{fig2}b. 
The answer, based on earlier\cite{sihy,sibert,husihy,stuma2,davhel}
works, should involve a nonlinear resonance zone in the phase space.
Indeed
the Poincar\'{e} surface of section
at $E=E^{0}_{20} \approx 18641$ cm$^{-1}$ 
shown in Fig.~\ref{fig4} indicates
a resonance zone juxtaposed between the zeroth-order
states as seen in the phase space. 
The surface of section is 
consistent with the state space view in Fig.~\ref{fig2}a
in that the primary 2:1 resonances do not appear
in the phase space. The 2:1 resonance zones, if present, would
show up in Fig.~\ref{fig2}a around $K_{1}/2 \approx \pm 2.6$. 
Hence a direct involvement of the 2:1 resonances is ruled out
but surely 
the existence of the 2:1s must be giving rise to a dynamical
barrier, in the form of the observed resonance
zone, mediating the tunneling.
In order to confirm that this nonlinear resonance is mediating the
dynamical tuneling it is necessary to extract an explicit
expression for the dynamical barrier and obtain the splitting.
For this purpose we apply methods of nonlinear dynamics\cite{joy,licht} 
on the classical Hamiltonian eq.4 with $g=0=\gamma$ 
and $\beta \neq 0$. The analysis below is not new and can be found in 
many of the earlier works 
including Ozorio de Almeida\cite{ozo} and Sibert\cite{sibert}.
For completeness sake only essential results are explicitly given
and technical aspects of classical perturbation theory will not
be discussed. 

To start with a canonical transformation is performed
$({\bf I},{\bm \theta}) \rightarrow ({\bf J},{\bm \psi})$ using the
generating function
\begin{equation}
F=(\theta_{1}-2\theta_{b})J_{1}+(\theta_{2}-2\theta_{b})J_{2}+\theta_{b}N
\end{equation}
The choice of the generating function implies 
$\psi_{k}=\theta_{k}-2\theta_{b}$, $J_{k}=I_{k}$ for $k=1,2$ and
$\psi_{3}=\theta_{b},N=2(I_{1}+I_{2})+I_{b}$. The resulting
Hamiltonian
\begin{eqnarray}
H({\bf J},{\bm \psi};N)&=&H_{0}({\bf J};N) +\epsilon
\beta_{c}(N-2J_{1}-2J_{2}) \\
&\times& [\sqrt{J_{1}}\cos\psi_{1}+\sqrt{J_{2}}\cos\psi_{2}] \nonumber
\label{cham2}
\end{eqnarray}
is ignorable in the angle $\psi_{3}$ implying the conserved quantity
$N=P+5/2$ which is the classical analog of the polyad number $P$. 
The above two-dimensional Hamiltonian contains the 
2:1 resonances and hence nonintegrable. However the phase space
from Fig.~\ref{fig4} suggests that at
$E \approx E^{0}_{2,0,12}$ the 2:1s do not have a direct role to play.
Another way of stating this is that the nonlinear frequencies
$\Omega_{k} \equiv \partial H_{0}({\bf I})/\partial I_{k}$ are
far away from the condition $\Omega_{1,2}-2\Omega_{b}=0$ for
the states in consideration. For the system parameters in this
study the $|2,0,8\rangle$ state is detuned by almost $800$ cm$^{-1}$.
Consequently 
a formal parameter $\epsilon$ has been introduced with the
aim of perturbatively removing the 2:1 resonances, characterized by
$\psi_{1,2}$ in eq.~\ref{cham2}, to $O(\epsilon)$. 
This can be done by invoking
the generating function 
\begin{equation}
G=\bar{J}_{1}\psi_{1}+
\bar{J}_{2}\psi_{2}+\epsilon[g_{1}\sin\psi_{1}+g_{2}\sin\psi_{2}]
\end{equation}
where the functions $g_{1,2}=g_{1,2}(\bar{J}_{1},\bar{J}_{2})$
are determined by the condition of the removal of the primary
2:1s to $O(\epsilon)$. 
The angles conjugate to ${\bar{\bf J}}$ are denoted by
${\bar{\bm \psi}}$. The new variables $({\bar{\bf J}},{\bar{\bm \psi}})$
are related to the variables $({\bf J},{\bm \psi})$ through the
relations $J_{1,2}=\partial G/\partial \psi_{1,2}$ and
$\bar{\psi}_{1,2}=\partial G/\partial {\bar J}_{1,2}$.
Using the identities:
\begin{eqnarray}
\cos(\epsilon z \sin(\varphi)) &=& {\cal J}_{0}(\epsilon z) + 2\sum_{l\geq 1}
{\cal J}_{2l}(\epsilon z) \cos(2l\varphi) \\
\sin(\epsilon z \sin(\varphi)) &=& 2\sum_{l\geq 0}
{\cal J}_{2l+1}(\epsilon z) \sin((2l+1)\varphi)
\end{eqnarray}
with ${\cal J}_{k}$ being the Bessel functions, it can be shown that
\begin{equation}
H({\bar{\bf J}},{\bar{\bm \psi}};N) = \bar{H}_{0}({\bar{\bf J}};N)+
\sum_{k}\epsilon^{k} H_{k}({\bar{\bf J}},{\bar{\bm \psi}};N) 
\label{cpertham}
\end{equation}
where
\begin{equation}
\bar{H}_{0}=\Omega_{s}(\bar{J}_{1}+\bar{J}_{2})+\alpha_{ss}(\bar{J}_{1}^{2}+
\bar{J}_{2}^{2}) + \alpha_{12}\bar{J}_{1}\bar{J}_{2}
\end{equation}
with $\Omega_{s}=\omega_{s}-2\omega_{b}+(x_{sb}-4x_{b})N,
\alpha_{ss}=x_{s}+4x_{b}-2x_{sb}$ and
$\alpha_{12}=8x_{b}-4x_{sb}+x_{ss}$.
At $O(\epsilon)$ 
\begin{equation}
H_{1}({\bar{\bf J}},{\bar{\bm \psi}};N) = 
\sum_{k=1,2}(\beta_{c} \bar{J}_{b} \sqrt{\bar{J}_{k}} +
g_{k}({\bar{\bf J}}) \widetilde{\Omega}_{k}) \cos \bar{\psi}_{k}
\end{equation}
where we have denoted $\bar{J}_{b} \equiv N-2(\bar{J}_{1}+\bar{J}_{2})$,
and $\widetilde{\Omega}_{k} \equiv \partial \bar{H}_{0}/\partial \bar{J}_{k}$.
The action $\bar{J}_{b}$ is essentially the number of bend quanta in
the system.
The choice of the functions
\begin{equation}
g_{k}({\bar{\bf J}})=-\beta_{c} \frac{\bar{J}_{b}
\sqrt{\bar{J}_{k}}}{\widetilde{\Omega}_{k}}
\end{equation}
eliminates the 2:1s to $O(\epsilon)$
and hence the $O(\epsilon^{2})$ term in the transformed
Hamiltonian eq.~\ref{cpertham} is determined to be:
\begin{equation}
H_{2}({\bar{\bf J}},{\bar{\bm \psi}};N)
= \sum_{k=1,2} \lambda_{k} (1+\cos 2\bar{\psi}_{k}) +
\lambda_{\pm} \cos(\bar{\psi}_{1} \pm \bar{\psi}_{2})
\label{cph2}
\end{equation}
The term $\cos(\bar{\psi}_{1}-\bar{\psi}_{2})$
is easily identified with a
1:1 resonance between the modes $1$ and $2$ and the strength
of this induced
1:1 resonance is obtained as 
\begin{equation}
\lambda_{-}({\bar{\bf J}};N)=\frac{1}{2}\beta_{c}^{2} 
\left[\frac{2(\widetilde{\Omega}_{1}+\widetilde{\Omega}_{2})+\alpha_{12}
\bar{J}_{b}}{\widetilde{\Omega}_{1}\widetilde{\Omega}_{2}}\right] \bar{J}_{b}
\sqrt{\bar{J}_{1}\bar{J}_{2}}
\label{dynb}
\end{equation}
Expressions for the strengths $\lambda_{1,2}$
and $\lambda_{+}$ are not needed for our analysis hereafter and hence
not explicitly given in this work.

At this stage the transformed Hamiltonian to $O(\epsilon^{2})$
in eq.~\ref{cph2} 
still depends on both the angles $\bar{\psi}_{1}$ and $\bar{\psi}_{2}$ and
hence non-integrable. In order to isolate the induced 1:1 resonance 
we perform a canonical transformation to the variables $({\bf K},{\bm \phi})$
using the generating function $G=(\bar{\psi}_{1}-\bar{\psi}_{2})K_{1}/2 +
(\bar{\psi}_{1}+\bar{\psi}_{2})K_{2}/2$ and average the resulting
Hamiltonian over the fast angle $\phi_{2}$. The resonance center,
$K_{1}^{r}=0$,
approximation is invoked resulting
in a pendulum Hamiltonian describing
the induced 1:1 resonance island structure seen in the surface
of section shown in Fig.~\ref{fig4}. Within the averaged approximation the
action $K_{2}=\bar{J}_{1}+\bar{J}_{2}$ is a constant of the motion
and can be identified as the 1:1 polyad associated with the secondary
resonance. 
The resulting integrable Hamiltonian is given by
\begin{equation}
\bar{H}(K_{1},\phi_{1};K_{2},N)=\frac{1}{2 M_{11}} K_{1}^{2}+
2g_{ind}(K_{2},N)\cos2\phi_{1}
\label{eq7}
\end{equation}
where
\begin{subequations}
\begin{eqnarray}
M_{11}&=&2(\alpha_{12}-2\alpha_{ss})^{-1} \\
g_{ind}(K_{2},N)&=&\frac{\beta_{c}^{2}}{2}\bar{f}(K_{2},N)
(N-2K_{2}) K_{2} \label{eq8}
\end{eqnarray}
with
\begin{equation}
\bar{f}(K_{2},N)=\frac{4(\Omega_{s}+\alpha_{ss}K_{2})+\alpha_{12}N}
{[2(\Omega_{s}+\alpha_{ss}K_{2})+\alpha_{12}K_{2}]^{2}}
\end{equation}
\end{subequations}
Note, and this is important for the discussion in section~\ref{psv1},
the averaging procedure {\em does not} remove the higher harmonics of
the induced 1:1. Thus by focusing on the induced 1:1 alone in
eq.~\ref{eq7} we have neglected all induced resonances of the form
$q:q$.
In terms of the zeroth-order quantum numbers $K_{1}=n_{1}-n_{2}$ and
$K_{2}=n_{1}+n_{2}+1 \equiv m+1$. Note that eq.~\ref{eq8}, more
generally eq.~\ref{dynb}, is an
analytic expression for the dynamical barrier seperating 
any two localized states in terms of the zeroth-order actions
$(I_{1},I_{2},I_{b})$. 

\begin{figure}
\includegraphics*[width=3in,height=3in]{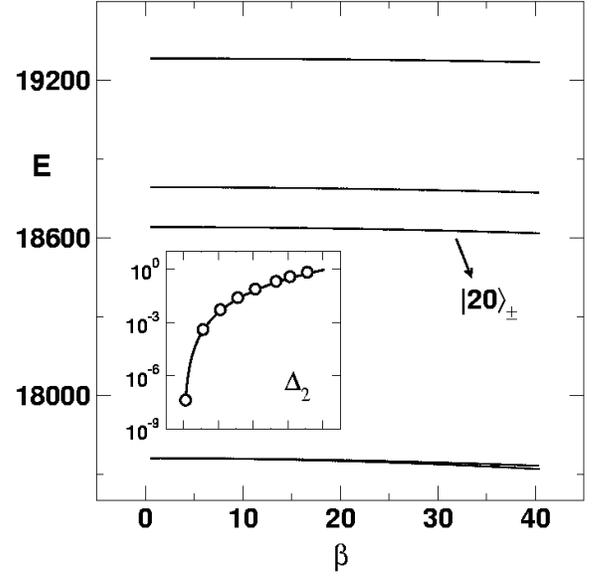}
\caption{Variation of the $|20\rangle_{\pm}$ doublet energies with
the 2:1 resonance strength $\beta$. Results correspond to $H_{A}$
and no exact or avoided crossings are observed. The inset
compares the splittings $\Delta_{2}$ calculated semiclassically
(open circles, eq.~\ref{indsp}) to the exact quantum (solid line)
over the same range of $\beta$ variation. Note that $\Delta_{2}$
increases by a factor of $10^{8}$ which is captured well by
the induced 1:1 resonance. All quantities are in cm$^{-1}$.}
\label{fig5}
\end{figure}

One can now use eq.~\ref{eq7} to calculate the
dynamical tunneling splitting 
of the degenerate modes $|n_{1}=r,n_{2}=0,n_{b}=2(P-r)\rangle$
and $|n_{1}=0,n_{2}=r,n_{b}=2(P-r)\rangle$ via\cite{stuma2}
\begin{eqnarray}
\frac{\Delta^{sc}_{r}}{2}
&=&g_{ind} \prod_{k=-(r-2)}^{(r-2)} \frac{g_{ind}}{E_{R}^{0}(r)-
E_{R}^{0}(k)}  \nonumber \\
&=& g_{ind} \frac{(M_{11} g_{ind})^{r-1}}{2^{r-1}[(r-1)!]^{2}}
\label{indsp}
\end{eqnarray}
where $E_{R}^{0}(k)=k^{2}/2M_{11}$ is the zeroth-order energy.
For our example with $r=2,m=2$ using the parameters of the Hamiltonian
we find $M_{11} \approx 1.32 \times 10^{-2}$ and $g_{ind} \approx 3.93$
cm$^{-1}$. Within the pendulum approximation the half-width of the resonance
zone can be estimated as $2\sqrt{2 M_{11} g_{ind}} \approx 0.64$. This
estimate, relalizing that $K_{1} \approx (I_{1}-I_{2})$, is in good
agreement with the surface of section in Fig.~\ref{fig4}.
The resulting splitting $\Delta_{2}^{sc} \approx 0.20$
cm$^{-1}$ agrees well with the exact splitting.
This proves that the induced 1:1 resonance arising from the
interaction of the two primary 2:1 resonances is mediating the dynamical
tunneling between the degenerate states. 
Moreover, from a superexchange perspective
it is illuminating to note that the splitting can be calculated
trivially by recognizing the secondary phase space structure (a viewpoint
emphasized in Ref.~\onlinecite{rat2} as well). 
This observation emphasizes
the superior nature of a phase space viewpoint on dynamical tunneling.
The analysis above also sheds light on the relation between
the quantum superexchange viewpoint and WKB approximation. 
Note that $g_{ind} \sim \beta^{2}$ and thus the phase space viewpoint,
with $\Delta_{2} \sim g_{ind}^{2}$ is equivalent to the
superexchange approach with $\Delta_{2} \sim \beta^{4}$. 
In this near-integrable situation this correspondence generalizes the
observation by Stuchebrukhov and Marcus about the equivalence of
superexchange and WKB methods for integrable systems\cite{stuma2}.

\subsection{Higher harmonics and level crossings}
\label{psv1}

In this section we address two issues relevant
in the context of dynamical tunneling.
The first issue has
to do with the role and relation of 
eigenvalue avoided crossings\cite{hel9599,ksjpca} to
nonlinear resonances and 
hence the phenomenon of dynamical tunneling.
Since we have just argued for the involvement of the 2:1 resonances
in mediating the dynamical tunneling between degenerate states
it is interesting to ask if the states are also involved in an
avoided crossing as the 2:1 strength $\beta$ is varied.
The answer to this question is negative
and Fig.~\ref{fig5} provides
evidence that the states $|20\rangle_{\pm}$
do not undergo any identifiable avoided or exact
crossings with varying $\beta$.
In fact the splittings (Fig.~\ref{fig5} inset)
predicted from eq.~\ref{indsp} agree fairly well
over the entire range of $\beta$.
It is also worthwhile noting that for the system $H_{A}$ 
the splittings $\Delta_{1,2,3}$ are
described quite well by a single induced 1:1 resonance. Therefore, in
general, it is not
necessary that dynamical tunneling between two
zeroth-order states $|n_{1}n_{2}\rangle$ and $|n_{1}+p,n_{2}-q\rangle$
be mediated by a $p:q$ nonlinear resonance alone.

A second issue has to do with the role of higher harmonics or Fourier
components in dynamical tunneling\cite{hel9599}. 
The doublet splitting $\Delta_{2}$
for $H_{A}$ was seen to arise from a induced 1:1 resonance. The
perturbative analysis, on the other hand, even in the averaged approximation
contains all the higher harmonics of the induced 1:1. These
Fourier components appear at higher
orders in $\epsilon$ and have very small, but finite, contribution
to $\Delta_{2}$. If for some reason the leading order induced 1:1
were to vanish then the higher Fourier components become important. 
For the present example the 2:2 component would be especially
significant. In order to demonstrate this effect consider
the Hamiltonian
\begin{equation}
\hat{H}=\hat{H}_{0}-|g|\hat{V}^{(12)}_{1:1}+\beta(\hat{V}^{(1b)}_{2:1}+
\hat{V}^{(2b)}_{2:1})
\label{add11}
\end{equation}
A 1:1 resonance has been added to $H_{A}$ with a strength opposite
in sign to the induced 1:1 resonance. The strength $g$ is varied for
fixed $\beta$ and there is a specific value of $g = g_{0}$ for which
the primary and the induced 1:1 resonance cancel each other out. 
The value of $g_{0}$ can be estimated by recognizing the
integrable pendulum approximation to the above Hamiltonian as
\begin{equation}
\bar{H} \approx \frac{1}{2 M_{11}} K_{1}^{2} +
(-|g| K_{2} + 2 g_{ind}) \cos 2\phi_{1}
\end{equation}
Thus around $|g|_{0} \approx 2g_{ind}/K_{2}$ the above leading
order approximation would predict a shutdown of dynamical tunneling.
From the estimates provided earlier it is easy to obtain
the value $|g|_{0} \approx 2.62$ cm$^{-1}$ for $\Delta_{2}$. 

\begin{figure}
\includegraphics*[width=3in,height=3in]{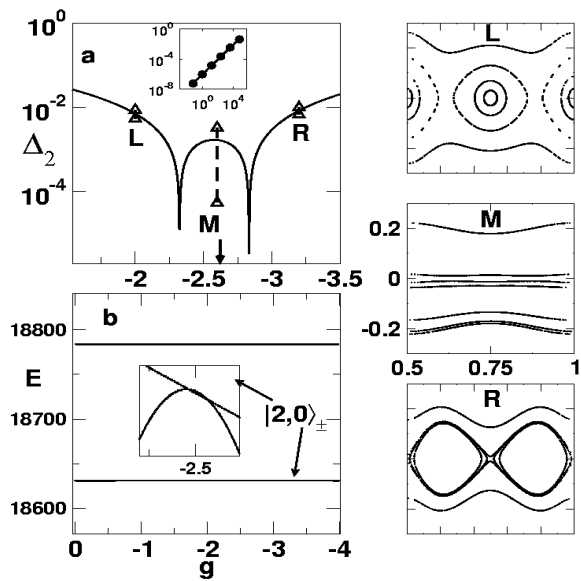}
\caption{(a) Variation of the exact $\Delta_{2}$ with the primary 1:1
strength $g$. Note the two dips and the relatively flat region
denoted by $M$. The arrow indicates the strength $g_{0}$. The
effect of adding a small 2:2 with strength $\gamma \approx \pm 4.1 \times
10^{-4}$ cm$^{-1}$ at specific places are shown as triangles.
Inset shows the linear scaling
of exact $\Delta_{2}$ (points)
with $\beta^{2}|g|_{0}$.
In (b) the variation of the relevant energy levels are shown. Inset
shows the details for the doublet $|20\rangle_{\pm}$ which undergo
two exact crossings. The three panels denoted by $L, M$, and $R$ show
the surface of sections corresponding to the regions labeled in (a).
The sections are computed using the full classical limit
Hamiltonian corresponding to $H_{A}-|g|V_{1:1}^{(12)}$.
Note the reorganization of the phase space in going from $L$ to $R$.}
\label{fig6}
\end{figure}

In Fig.~\ref{fig6}a the variation of the exact 
splitting $\Delta_{2}$ is shown
as a function of $g$. The results clearly show the destruction of
tunneling near $g_{0}$ but a more significant observation is that
at $g_{0}$, labeled by region $'M'$ in Fig.~\ref{fig6}a, the actual value
of the splitting is far from zero. This is in contrast to the
prediction of shutdown of dynamical tunneling based on the
integrable pendulum approximation above. The corresponding phase
space in fig.~\ref{fig6}M
indeed shows that the 1:1 resonance zone has almost vanished. So
what is mediating the dynamical tunneling in region $M$? 
Arguments based on some residual 1:1 coupling are quickly dispelled by
noting that the width of the resonance zone in Fig.~\ref{fig6}M is nowhere
close to the required value. Chances of a strong avoided crossing
leading to the enhanced splitting are ruled out by inspecting
Fig.~\ref{fig6}b. Perhaps there is a broad avoided crossing, as suggested
by the exact crossing of the doublets 
$|20\rangle_{\pm}$ in fig.~\ref{fig6}b (inset), 
leading to the two dips in $\Delta_{2}$ (cf. fig.~\ref{fig6}a), 
with the upper state which happens to be a normal mode state.
It is relevant to note that the energy seperation between the
doublet and normal mode state is almost equal to the mean level spacing.
Moreover a superexchange calculation of $\Delta_{2}$ at $M$ 
including the required minimal chains (cf. eq.~\ref{mhopt})
$\Gamma_{400},\Gamma_{020}$, and $\Gamma_{210}$,
yields a value $1.70 \times 10^{-3}$ cm$^{-1}$ which is in excellent
agreement with the exact value of $1.65 \times 10^{-3}$ cm$^{-1}$.
All this suggests that in the region $M$ a tiny, but leading order,
2:2 with strength 
$\gamma = \Delta_{2}/4 \approx 4.1 \times 10^{-4}$ cm$^{-1}$ is mediating
the dynamical tunneling. A few important observations provide further
support and we briefly outline them. Firstly the splitting in region
$M$ are very sensitive to adding a perturbation $\pm \gamma V^{(12)}_{2:2}$
to the Hamiltonian in eq.~\ref{add11} whereas the
regions $L$ and $R$ are far less sensitive as shown in Fig.~\ref{fig6}a.
Secondly, the classical perturbation analysis carried out to higher
orders reveals the emergence of an 
induced 2:2, scaling as $\beta^{2}g$ at $O(\epsilon^{3})$, with strength
comparable to $\gamma$. Indeed the exact $\Delta_{2}$ in region $M$
for different values of $\beta$ scales linearly with $\beta^{2}g$.
The superexchange calculation also shows that the 
contribution from the minimal family
$\Gamma_{210}$ {\it i.e.,} $\beta^{2}g$ nearly, but not exactly, balances
the contribution from $\beta^{4}$, and $g^{2}$ families. 
The analysis here is an example of what
Pearman and Gruebele\cite{pear} 
call as 'phase cancellation' effect in perturbative
chains. Evidently, the analog of this effect in phase space has to do with
the destruction of nonlinear resonances. Although only the $r=2$ case has
been shown here, similar effects arise for the higher excitations
and can be more complicated to interpret.

\subsection{Multiple resonances: Resonance-assisted tunneling}
\label{psv2}

\begin{figure}
\includegraphics*[width=3in,height=3in]{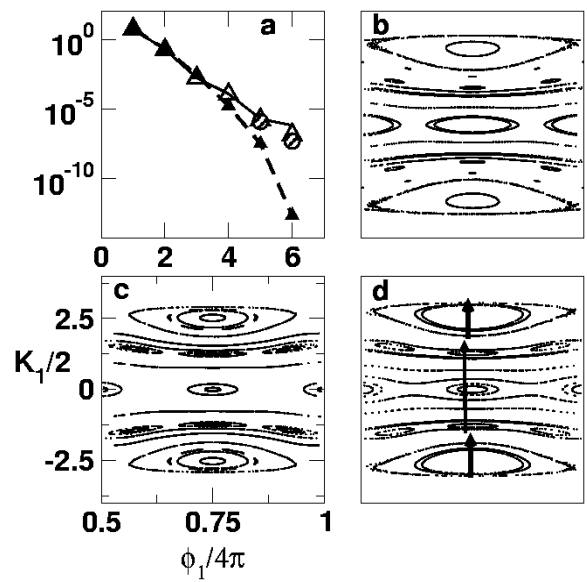}
\caption{(a) Comparison of the exact (open triangles),
superexchange (solid line), and the
phase space resonance based (filled triangles)
splittings for $H_{A}$ versus excitation quanta in the OH-stretch.
Multiple resonance corrections to $r=5,6$ are shown by shaded circles.
See text for details.
In (b), (c), and (d) the surface of
sections at energies
corresponding to $r=4,5$, and $6$ respectively are shown.
The axes
ranges for (b) and (d) are identical to those shown in (c).
In (d)
one possible dynamical tunneling sequence is shown by arrows.}
\label{fig7}
\end{figure}

One of the key lessons learnt from the preceeding discussion is that
nonlinear resonances have to be identified in order to
obtain a clear phase space picture of dynamical tunneling. 
Given the near-integrable phase space of $H_{A}$ it is tempting
to apply eq.~\ref{indsp} to calculate the splitting of the other
doublets corresponding to increasing stretch excitations. In particular
the monotonic decrease with increasing stretch excitations, seen
in Fig.~\ref{fig3}, seems to fit well with the notion of a single
resonance (the induced 1:1 here) mediating the tunneling. 
The naivety of such a viewpoint is illustrated in Fig.~\ref{fig7}a.
Clearly for $r=1,2,3$ the expectation holds but deviations arise
already at $r=4$ and the splittings are many orders of magnitude smaller
than the quantum results for $r=5,6$. 
The reason for this large error is immediately clear from the surface
of sections shown in Fig.~\ref{fig7}b,c,d. The primary 2:1 resonance
has appeared in the phase space and coexists with the induced 1:1.
The states corresponding to $r=4,5$ are 'resonant' local modes and
a Husimi representation 
of the eigenstate in the classical
phase space (cf. Fig.~\ref{fig10}a) confirms their nature. 
The $r=5$ doublets are in fact localized in the large 2:1 resonance islands.
This can also be anticipated from the state space location of the
resonance zones shown in Fig.~\ref{fig2}a. Thus a single rotor integrable
approximation is insufficient to describe the dynamical tunneling for
such cases. Notice the rich structure of the phase spaces in
Fig.~\ref{fig7}b,c,d resulting from the presence of two 'distant'
and 'nonoverlapping' (in the Chirikov\cite{chiri} sense) resonances - a clear
manifestation of the nontrivial nature
of the near-integrable systems. 

Is it possible to use the phase space structure, for example in
Fig.~\ref{fig7}d, and calculate the splitting $\Delta_{6}$?
The phase space shows the existence of the primary 2:1s, the
induced 1:1, and a thin multimode resonance. For now we will
ignore the multimode resonance and consider the primary 2:1s and the
induced 1:1 to be independent. This is a strange approximation
given that both the induced 1:1 and the multimode arise due to the 2:1s.
However, since they are 'isolated' from each other and lacking
any other simple approach we take this viewpoint and compute the 
splitting. This approach has been advocated earlier
by Brodier, Schlagheck and Ullmo in their analysis\cite{rat2} 
of resonance-assisted tunneling in kicked
maps. In the present case the effective $\hbar$ is too large
to be in the semiclassical limit and hence quantitative
accuracies are not expected. At the same time this is a blessing in
disguise since we do not have to deal with the the myriad of resonance
zones in the phase space!
Schematically we imagine the following (cf. Fig.~\ref{fig7}d):
\begin{equation}
|60\rangle \stackrel{\beta_{eff}}{\longrightarrow} |40\rangle 
\stackrel{g_{ind}}{\longrightarrow} |04\rangle
\stackrel{\beta_{eff}}{\longrightarrow} |06\rangle
\end{equation}
where $\beta_{eff}$ is the effective coupling
across the 2:1 islands and $g_{ind}$ is the effective 1:1 coupling
derived in the previous section. 
It is important to note that the $g_{ind}$ value appropriate
to $r=6$ needs to be used {\it i.e.,} the resonance zone
width as seen in the surface of section in Fig.~\ref{fig6}d.
The effective 2:1 couplings can
be extracted by approximating the 2:1 resonances, for instance
the top island in Fig.~\ref{fig7}d, by a pendulum
Hamiltonian:
\begin{equation}
\frac{1}{2M_{21}}(J_{1}-J_{1}^{r})^{2}+2\beta_{p} \cos\psi_{1}
\end{equation}
with $M_{21} = (2\alpha_{ss})^{-1}$ and $\beta_{p} = \beta_{c}
(N-2J_{2}-2J_{1}^{r})\sqrt{J_{1}^{r}}/2$. 
The resonance center $J_{1}^{r} = -M_{21}(2\Omega_{s}+\alpha_{12})/2
\approx 5.7$. The effective coupling is thus estimated as:
\begin{equation}
\beta_{eff} \approx \frac{\beta_{p}^{2}}{E_{6}^{0}-E_{5}^{0}}
\end{equation}
with $E_{n}^{0} = (\hbar(n_{1}+1/2)-J_{1}^{r})^{2}/(2M_{21})$.
The splitting is then calculated via the expression
\begin{equation}
\Delta_{6} \approx \left(\frac{\beta_{eff}}{E_{6}^{0}-E_{4}^{0}}\right)
\times \Delta_{4}^{sc} \times 
\left(\frac{\beta_{eff}}{E_{6}^{0}-E_{4}^{0}}\right)
\end{equation}
where $\Delta_{4}^{sc}$ is the splitting for $r=4$ calculated
assuming the induced 1:1 resonance alone. Using the numerical values
for the various parameters we find $\Delta_{6} \approx 4.8 \times 10^{-8}$
cm$^{-1}$ as compared to the exact value of $1.4 \times 10^{-7}$ cm$^{-1}$.
This simple estimate is quite encouraging if one also considers the
fact that the induced 1:1 alone would give $\sim 10^{-13}$ cm$^{-1}$ and
the superexchange value is $6.5 \times 10^{-7}$ cm$^{-1}$.
Similar calculation for $r=5$ doublet also improves the result
as shown in Fig.~\ref{fig7}a. This simple minded scheme has been tested
for other parameters and the results are comparable to the
present case. Further support for such an approach comes from considering
the $r=6$ doublet splitting with an additonal 2:2 resonance present {\it i.e.,}
described by the Hamiltonian $H_{B}$. From Fig.~\ref{fig3} it is clear
that the addition of the 2:2 results in about four orders of
magnitude increase in $\Delta_{6}$. The phase space is still near-integrable
and a calculation based on the 2:2 alone underestimates the exact result
of $3.7 \times 10^{-3}$ cm$^{-1}$ by nearly a factor of four. Once again
employing the approach of hopping across the 2:1 using $\beta_{eff}$, 
followed by the 2:2 
connecting $|40\rangle$ to $|04\rangle$, and hopping across the
symmetric 2:1 island improves the result yielding a splitting of
about $2.8 \times 10^{-3}$ cm$^{-1}$. 
Interestingly the superexchange
calculation has a dominant contribution of similar
magnitude from the family $\Gamma_{402}$
which is readily identified with the $\beta^{4}\gamma^{2}$ 'path'
in phase space outlined above. The analysis here indicates that it
is conceivable that nonlinear resonances can lead to dynamical
tunneling between energetically similar regions of phase space even
in nonsymmetric situations. 

\begin{figure}
\includegraphics*[width=3in,height=2in]{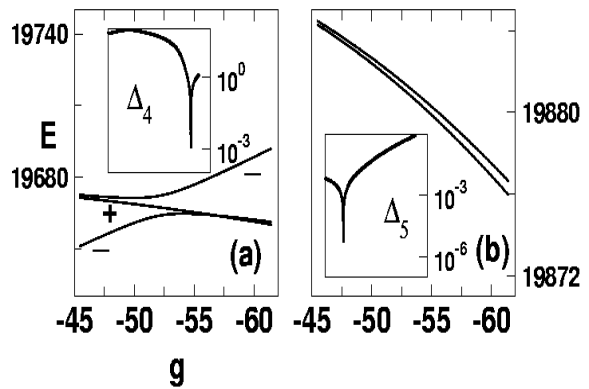}
\caption{Variation of eigenenergies with the 1:1 strength parameter $g$.
(a) $r=4$ doublet involved in an avoided crossing with a $-$ parity
normal mode state. Inset shows the behaviour of $\Delta_{4}$ with varying $g$.
(b) Same as in (a) for the $r=5$ doublets. All parameters are in cm$^{-1}$.
Note that the inset x-axis range for $g$ is from $0$ to
$-60$ cm$^{-1}$. Chaos is visible in the phase space for $|g| > 30$ cm$^{-1}$.}
\label{fig8}
\end{figure}

\subsection{Mixed phase space: Chaos-assisted tunneling?}
\label{psv3}
Up until this point dynamical tunneling was investigated with
the underlying phase space being integrable or near-integrable. We now
look at the full system $H_{C}$ which exhibits mixed chaotic-regular
phase space. The phenomenon of chaos-assisted (suppressed) tunneling (CAT)
has to do with the coupling of quantum states, localized on two
symmetry-related regular regions of phase space, with one or more
'irregular' states delocalized over the chaotic sea\cite{cat}. 
Clearly
such processes cannot be understood within a two-state picture
and the sensitivity of the nature of the chaotic
states to parametric variations is manifested in the splitting
fluctutations of the tunneling doublets. It is also reasonable to
expect that the effective $\hbar$ in the system needs to be sufficiently
small in order for CAT to manifest itself. 
Perhaps it is useful to note a recent analysis\cite{mou} by Mouchet and
Delande wherein they argue that even in cold atom tunneling
experiments the effective $\hbar$ is not small enough to observe CAT.
The system investigated herein is nowhere close to such limits
and at the outset we do not expect to see CAT.
In other words for our full system described by $H_{C}$ only 
resonance-assisted tunneling (RAT) mechanisms should be sufficient to
explain the relevant splittings\cite{mou}. 

\begin{figure}
\includegraphics*[width=3in,height=3in]{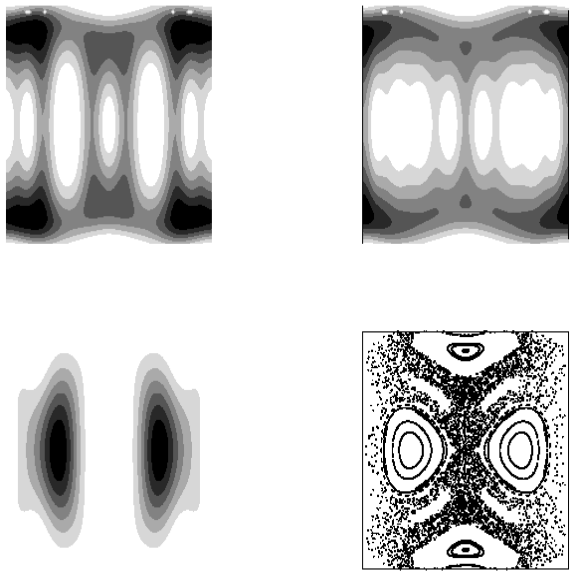}
\caption{Husimi distributions of the $r=4$ doublets and a close
lying normal mode state for the full system $H_{C}$.
The axes ranges are identical to those
shown in fig.~\ref{fig7}b.
The dark regions correspond to large values while the
light regions signify smaller values.
The top row shows the Husimis for
the $-$ and $+$ parity states while the bottom left is for the
$-$ parity normal mode state. The bottom right panel shows
the phase space with clearly identified 1:1 and 2:1 resonance islands.
Note the delocalized Husimis of the doublets in contrast to
that of the normal mode state.}
\label{fig9}
\end{figure}

We focus on the doublets $r=4,5,6$ since the phase
space shows substantial chaos at the corresponding energies. In
Fig.~\ref{fig3} it is seen that the $r=4$ case showed marked
deviations from monotonicity. Moreover the superexchange calculation
predicted a much higher $\Delta_{4}$. In Fig.~\ref{fig8}a we show the
variation of energy levels with the 1:1 strength $g$ and a clear
avoided crossing between one of the tunnel doublets (with $-$ parity)
and a normal mode state is observed. The inset in Fig.~\ref{fig8}a
implicates the avoided crossing in the observed exact crossing between
the tunnel doublets themselves leading to the suppression of tunneling.
In Fig.~\ref{fig9} the phase space 
Husimi\cite{husimi} representations (refer to \onlinecite{ksjcp} for
details regarding the computation of the Husimi distributions) of the three
states are shown with the corresponding phase space. Notice that the
normal mode ('intruder') state has a much more regular Husimi in the
phase space as compared to the doublet Husimis themselves. In particular
the $-$ doublet is extensively delocalized over the phase space. 
Given the value of the
effective $\hbar$ it is not possible to
declare the $-$ parity doublet
to be a chaotic state. Note however that the doublets corresponding
to $r=4$ are essentially living in the chaotic regions. For the near-integrable
subsystems $H_{A}$ and $H_{B}$ the $r=4$ doublets live on the
separatrix associated with the 2:1 resonance. So this is an instance
where a regular intruder state is affecting the tunneling between
irregular doublets. No attempt is made here to calculate the tunneling
splitting semiclassically. However we note that the exact value 
$\Delta_{4} \approx 0.1$ cm$^{-1}$ as compared to
the superexchange value of $2.3$ cm$^{-1}$. Interestingly with
further increase of $g$, resulting in more chaos, the superexchange
calculation is only a factor of two higher than the exact value.

\begin{figure}
\includegraphics*[width=3in,height=3in]{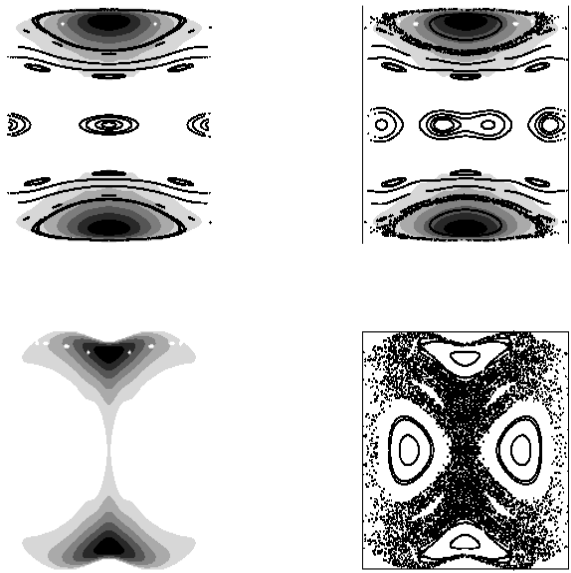}
\caption{The top panels show the Husimis superposed on
phase space for the $+$ symmetric partner of the $r=5$ doublet
in the near integrable cases.
The axes ranges are identical to those shown in fig.~\ref{fig7}c.
Top left is for $H_{A}$ (2:1 only) and
top right is for $H_{B}=H_{A}+2:2$. The Husimis in both instances
are localized in the 2:1 islands. The bottom left panel shows the
Husimi for the full case $H_{C}$ and the corresponding phase space
is shown in the bottom right. Note the small isthumus across the
chaotic sea connecting the two islands.}
\label{fig10}
\end{figure}

On the other hand in Fig.~\ref{fig8}b the variation of the $r=5$
doublet energies are shown as a function of the 1:1 parameter $g$. This
case does not show any avoided crossing. The Husimis for this doublet
corresponding to the systems $H_{A}, H_{B}$ (near-integrable) and
$H_{C}$ (mixed) are shown in Fig.~\ref{fig10}a,b, and c respectively.
It is notable that the splitting in going from $H_{A}$ to $H_{B}$ 
increases by more than three orders of magnitude (cf. Fig.~\ref{fig3}). 
The Husimis and
surface of sections in Fig.~\ref{fig10}a,b are quite similar except
for the 2:2 resonance zone clearly visible in Fig.~\ref{fig10}b. However,
the 2:2 alone cannot mediate dynamical tunneling between the localized
states in the 2:1 zone. As in the previous section, excitation out to
the seperatrix state $|40\rangle$ followed by tunneling across the
2:2 zone to $|04\rangle$ and finally coupling to the symmetric
$|05\rangle$ state captures quantitatively the tunneling enhancement.
This is a clear manifestation of RAT in the system. Adding the primary
1:1 resonance, resulting in $H_{C}$, the doublet splitting is further
enhanced by two orders of magnitude (cf.Fig.~\ref{fig3}). 
The phase space as shown in
Fig.~\ref{fig10} is clearly of a mixed nature. The large 1:1 resonance
zones are visible and tentatively one assigns the enhancement in splitting
to the 1:1 coupling. The 1:1 alone
can account for one order of magnitude enhancement.
However of all the possible resonant subsystems
considered, quantum mechanically or semiclassically, the best estimate is
a factor of two too small as compared to the $\Delta_{5}$ for the
full system. In this context an inspection of Fig.~\ref{fig10} Husimi
for the full case reveals amplitude spreading through the classically
chaotic region. This is in stark contrast to the well localized Husimis
in the near-integrable cases. Again the effective $\hbar$ is too large
to implicate the classical stochasticity for part of the
enhancement in $\Delta_{5}$. 
Although the case for $r=6$ is not shown here $\Delta_{6}$
is enhanced (cf. fig.~\ref{fig3}) by
four orders of magnitude in going from $H_{A}$ to $H_{B}$. This
enhancement
could again be captured quantitatively by the semiclassical analysis. However
the best integrable estimate based on the 1:1 and the 2:2 resonances is
almost an order of magnitude too small for the full system $H_{C}$. 

The upshot of the precceding analysis is that no integrable
subsystem can account for the splittings of $H_{C}$ and one atleast needs
some chaos to be present. Even subsystems with chaos, for instance
2:1s $+$ 1:1, on exact diagonalization yield splttings which
are about a factor of two in error. This points to a fairly subtle
involvement by all the resonances, irrespective of their strengths,
and is not understood well at this point of time. 
Reducing the effective $\hbar$ in the system is a theoretical tool
to clarify the picture.
For our system, a nonscaling one,
it is not very easy to perform $\hbar$-scaling computations. Nevertheless
some preliminary calculations indicate that the near-integrable cases
show exponential fall off with $\hbar^{-1}$ whereas the
mixed cases exhibit fluctuations. In particular the $r=4$ doublet splitting
exhibits an algebraic dependence $\hbar^{-\alpha}$ rather 
than the usual 
$\exp(-1/\hbar)$ dependence.

\section{Conclusions}
\label{summ}
In this work 
an intimate connection between dynamical tunneling
and the resonance structure of the classical phase space is established. 
The order and widths of the resonances determine the explicit 
dynamical barriers in the zeroth-order action (quantum number) space.
The current study indicates that it is safer to 'blame' dynamical
tunneling on the nonlinear resonances without invoking further
connections between avoided crossings and the resonances.
If accidental avoided crossings occur, as they do in near-integrable, mixed and
chaotic systems, then dynamical tunneling can be enhanced or suppressed
further.
This study also lends support to earlier suggestions\cite{hel9599} that
even very small resonant couplings can be the cause of the
experimentally observed narrow spectral
clusters. The resonances responsible for such spectral clusters can
be primary or induced resonances and with small but significant amount
of stochasticity in the phase space the consequences for energy flow
can be nontrivial. 

In near-integrable situations, generic to lower energy regimes in
large molecules, a combination of more than one resonance zone
can control dynamical tunneling. Not only is it important to identify the
relevant resonance zones but it is also necessary to prescribe
a route to calculate the splittings. 
It is significant to note that neither
of these problems are easy to solve. Again, and not surprisingly, the
lack of understanding of the structure of multidimensional phase space
is the main bottleneck. In this regard the superexchange approach
works well but such an accuracy for high dimensional systems, with
mixed regular-chaotic phase spaces, is not
guaranteed. As shown in the present work and in a previous work\cite{self}
the superexchange approach is prone to errors, atleast by an order
of magnitude, whenever states are involved in avoided crossings. It
remains to be seen if going beyond the minimal paths prescription, with
some clever 'resummation' tricks, would
lead to quantitative improvements.  
We have shown that treating the resonance zones as
pendulums and coupling across them to connect the two nearly
degenerate states provides an approximate route to calculating
the tunnel splittings. This chain of resonance zones, in some sense, provides
a semiclassical basis for the success of the vibrational superexchange
approach.  
The current work has shown the close correspondence
between various terms contributing in the superexchange approach
and the underlying resonances in the phase space.
The superexchange mechanism itself has its roots in the
field of electron transfer in molecular systems\cite{nitz}.
In this context it is interesting to 
note the analogy between the role played by the 'bridge' states
connecting the donor and acceptor sites for 
electron transfer\cite{nitz} and the
resonance zones connecting two symmetry related, distant states
localized in phase space. Some similarities, perhaps formal, 
to this work showing
the promotion of dynamical tunneling by multiple resonances
and a recent work\cite{uri}
demonstrating promotion of deep tunneling through molecular
barriers by electron-nuclear coupling further exemplify the
close parallels between the phenomena of off-resonant
electron transfer and dynamical tunneling.
Regarding the various mechanisms of dynamical tunneling in molecules
this work suggests that for small molecules with low density of states (large
effective $\hbar$) the dominant mechanism would be resonance-assisted.
For molecules with sufficiently high density of states (sufficiently small
effective $\hbar$) both chaos and resonances can mediate dynamical tunneling
and the dominant mechanism is decided by the energy of
the doublets and the location of the
doublets in the corresponding phase space.

Although the analysis was on coupled three-mode systems the conclusion
remains valid in general for systems which exhibit local mode behaviour.
This is based on our analysis, along similar lines, for many other
systems which are described more naturally by 
local (D$_{2}$O, H$_{2}$S) or normal mode (SO$_{2}$) limits.
For systems with small or vanishing anharmonicities, where
the issue of the existence of localized modes itself
is moot, the various
resonances are still important but any calculation
based on a pendulum
approximation is clearly invalid.   
Thus spectra of
molecules involving light atom stretch-bend modes are good candidates
to observe the fingerprints of dynamical tunneling. In particular we
suggest the high overtone spectral regions of the water molecule
as one possibility. 
This suggestion is tentative since a model Hamiltonian has been
utilized for the analysis and it is probable that the polyad picture
will break down at such high energies. One possibility is to do a detailed 
analysis on the best potential energy surface available and/or explicitly
break the polyad by adding further weak resonances. Both approaches
result in genuine three degree of freedom phase spaces and thus 
a straightforward extension is not easy. 
Apart from the limitation in visualizing the global phase space
structures this has to do with
the fact that there are classical phenomena\cite{licht}
in three or more degrees of freedom
that could modify certain aspects of the lower dimensional
analysis. For instance in three or higher degree of freedom systems two
long time phenomena are controlled by nonlinear 
resonances - Arnol'd diffusion\cite{licht,commardif}
and dynamical tunneling. Precious little is known about either 
of these phenomena
and their possible spectral manifestations in such situations and needs
further study\cite{compet}.
As a final note we comment on the possibility of observing long lived
local excitations in systems with multiple resonances. 
This work shows
that additional resonances can modify the near-degeneracies via
CAT and RAT. But this modification, by the same additional
resonances and resulting chaos,
can go either way due to multistate interactions! 
Moreover, as shown in this work, additional resonances
and presumably coupling to rotations\cite{lehkk,orti}
can also conspire to increase the degeneracy.
Coherent manipulation of tunneling using external
fields has been demonstrated before\cite{mfield,cohdesfield} 
and there are reasons to think
that such ideas can be utilized in the present context as well.

\section{Acknowledgements}

It is a pleasure to thank Peter Schlagheck for critical 
and illuminating discussions.
I am grateful to Prof. Klaus Richter 
for the hospitality and support at the Universit\"{a}t Regensburg where
part of this work was done.


\begin{thebibliography}{}
\bibitem{tenpers}{The experimental value for the splitting of the
$(40)^{\pm}4$ pair is about $1.0$ cm$^{-1}$. J. Tennyson, personal
communication.}
\end{thebibliography}
\end{document}